\documentclass[aps,twocolumn,showpacs,superscriptaddress,floatfix]{revtex4-1}
\usepackage[utf8]{inputenc}
\usepackage{graphicx}
\usepackage{amsmath}
\usepackage{amsfonts}
\usepackage{amssymb}
\usepackage{mathtools}
\usepackage{braket}
\usepackage[breaklinks=true,colorlinks,citecolor=red]{hyperref}
\usepackage[english]{babel}

\usepackage[all]{xy}

\DeclarePairedDelimiter\floor{\lfloor}{\rfloor}
\DeclarePairedDelimiter\expect{\langle}{\rangle}

\newcommand{\cf}{{\itshape cf.\ }}
\newcommand{\eg}{{\itshape e.g.\ }}
\newcommand{\ie}{{\itshape i.e.\ }}

\newcommand{\tr}{\operatorname{tr}}
\newcommand{\coloneq}{\mathrel{\mathop:}=}

\DeclarePairedDelimiter\abs{\lvert}{\rvert}%
\makeatletter
\let\oldabs\abs
\def\abs{\@ifstar{\oldabs}{\oldabs*}}
\makeatother

\newcommand{\DC}{\Delta_\mathrm{c}}

\newcommand{\omrec}{\omega_\mathrm{R}}
\newcommand{\nmax}{n_\mathrm{c}}
\newcommand{\deltak}{\Delta_k}
\newcommand{\rhostationary}{\rho_{\mathrm{ss}}}
\newcommand{\rhoalpha}{\rho_{\alpha}}
\newcommand{\kfermi}{k_{\mathrm{F}}}
\newcommand{\efermi}{\epsilon_{\mathrm{F}}}
\newcommand{\kB}{k_{\mathrm{B}}}
\newcommand{\etac}{\eta_{\mathrm{c}}}

\begin{document}

\title{Self-ordered stationary states of driven quantum degenerate gases\\in optical resonators}

\author{Raimar M.~Sandner}
\affiliation{Institut f{\"u}r Theoretische Physik, Universit{\"a}t Innsbruck, Technikerstra{\ss}e~25, A-6020~Innsbruck, Austria}

\author{Wolfgang Niedenzu}
\affiliation{Institut f{\"u}r Theoretische Physik, Universit{\"a}t Innsbruck, Technikerstra{\ss}e~25, A-6020~Innsbruck, Austria}
\affiliation{Department of Chemical Physics, Weizmann Institute of Science, Rehovot~7610001, Israel} 

\author{Francesco Piazza}
\affiliation{Institut f{\"u}r Theoretische Physik, Universit{\"a}t Innsbruck, Technikerstra{\ss}e~25, A-6020~Innsbruck, Austria}

\author{Helmut Ritsch}
\email{Helmut.Ritsch@uibk.ac.at}
\affiliation{Institut f{\"u}r Theoretische Physik, Universit{\"a}t Innsbruck, Technikerstra{\ss}e~25, A-6020~Innsbruck, Austria}

\begin{abstract}

We study the role of quantum statistics in the self-ordering of ultracold bosons and fermions moving inside an optical resonator with transverse coherent pumping. For few particles we numerically compute the nonequilibrium dynamics of the density matrix towards the self-ordered stationary state of the coupled atom-cavity system. We include quantum fluctuations of the particles and the cavity field. These fluctuations in conjunction with cavity cooling determine the stationary distribution of the particles, which exhibits a transition from a homogeneous to a spatially ordered phase with the appearance of a superradiant scattering peak in the cavity output spectrum. At the same time the cavity field $Q$-function changes from a single to a double peaked distribution. While the ordering threshold is generally lower for bosons, we confirm the recently predicted zero pump strength threshold for superradiant scattering for fermions when the cavity photon momentum coincides with twice the Fermi momentum. 
\end{abstract}

\date{\today}

\maketitle

\section{Introduction}

Ultracold particles moving in an optical resonator are a fast growing research field since the advent of laser cooling~\cite{Ritsch2013,Mekhov2012Quantum,stamper2014cavity}. The establishment of cavity cooling~\cite{Horak1997} extended cooling possibilities to a large class of linearly polarisable particles and in the last years cavity optomechanics~\cite{Aspelmeyer2014cavity} has experienced a tremendous boost, both theoretically and experimentally~\cite{stamper2014cavity}. Recently, cavity cooling was even extended to the subrecoil regime in the domain of ultracold quantum gases where the nonlinear coupled dynamics of field and particles require a full quantum description~\cite{Wolke2012,Sandner2013subrecoil,Elliott2015Probing}. 

\par

\begin{figure}[b]%
  \includegraphics[width=0.6\columnwidth]{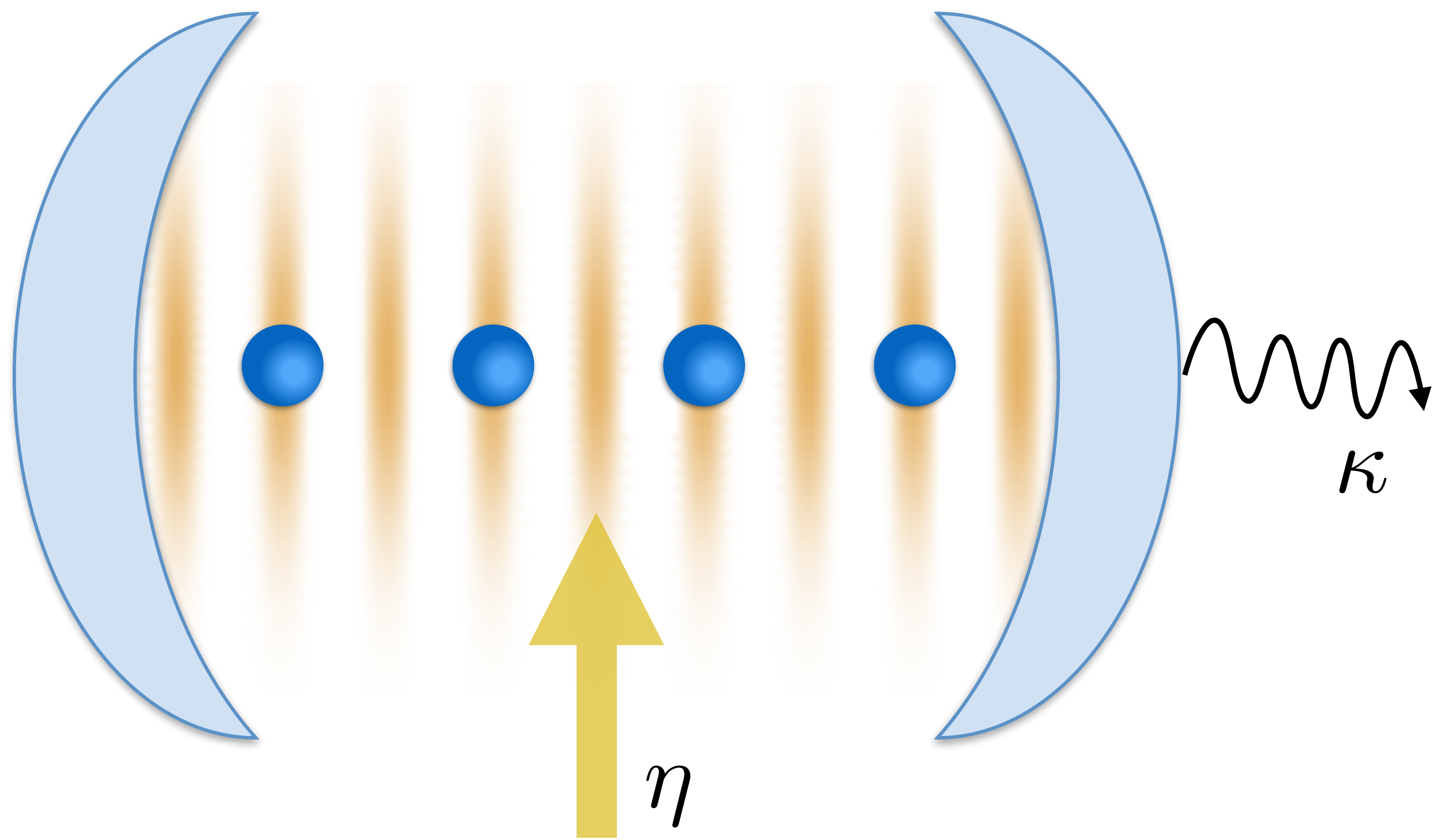}%
  \centering%
  \caption{Polarisable particles moving along the axis of a standing-wave optical resonator are coherently pumped in the transversal direction. Above a critical threshold the particles collectively scatter photons into the cavity while organising themselves in the emerging optical potential.}\label{fig:system}%
\end{figure}%
Particularly rich physical dynamics appear for transversally illuminated particles inside high-$Q$ optical resonators (\cf fig.~\ref{fig:system}), where the cavity-induced backaction of the scattered field onto the particles induces a phase transition to crystalline particle order. First described semiclassically~\cite{domokos2002collective, Asboth2005selforganization,Griesser2010,Niedenzu2011kinetic,Griesser2012cooperative, Schuetz2013cooling,Schuetz2014Prethermalization} this phenomenon also appears quantum-mechanically as a phase transition at zero temperature~\cite{Vukics2007a,Fernandez2010quantum, Konya2011,Habibian2013bose, Li2013Lattice,Piazza2013bose,Bakhtiari2015Nonequilibrium}. Above a critical threshold pump laser intensity the atoms form self-sustained regular Bragg lattices, which maximise the collective scattering of laser light into the resonator (superradiance). 
Self-organisation has been experimentally demonstrated with both thermal gases~\cite{Black2003observation,Arnold2012self} and atomic Bose-Einstein condensates (BECs)~\cite{Baumann2010dicke,Mottl2012Roton,Kessler2014Steering}.
It bears a close connection to the generic model of the Dicke superradiant phase transition~\cite{Dicke1954coherence,Hepp1973superradiant,Hioe1973Phase}.

\par

Interestingly, the quantum statistical properties of the particles have a decisive influence onto the self-ordering dynamics. As a particularly striking effect, a resonant reduction of the superradiance threshold for fermions almost towards zero pump amplitude has been predicted when the cavity photon momentum coincides with twice the Fermi momentum~\cite{Piazza2014Umklapp, Keeling2014Fermionic,Chen2014Superradiance}. 

\par

Since we are dealing with an open system with atom pumping and photon leakage through the mirrors, fluctuations and nonthermal equilibrium phenomena are essential parts of the full dynamics, as already observed experimentally~\cite{Brennecke2013real, Klinder2015Dynamical}.
We consider here the so-called dispersive limit at high laser to atom detuning, where the particles constitute a dynamic refractive index within the resonator and only coherently scatter pump laser light into the cavity. The relative phase and magnitude of the scattered photons depend on the particle positions, while the motion of the latter is governed by the optical dipole force exerted by the cavity and pump fields~\cite{Ritsch2013}. Recent experiments~\cite{Wolke2012,Kessler2014Steering, Klinder2015Dynamical} have achieved cavity loss rates small enough to be comparable to the particle recoil energy scale. Lacking separate timescales, the light field then does not adiabatically follow the particle motion, inducing an even more complex correlated or even entangled particle-field dynamics. 

\par

At the semiclassical level and for distinguishable particles the nonthermal properties of the corresponding stationary equilibrium state and its transient dynamics have been studied recently~\cite{Niedenzu2011kinetic,Griesser2012cooperative,Schuetz2013cooling,Schuetz2014Prethermalization}. A related study was carried out for bosons at zero temperature with emphasis on the effect of cavity losses on the depletion of a BEC~\cite{Konya2012finite}. In addition, the effect of fermionic statistics on the nonequilibrium stationary state has been explored with perturbative diagramatic techniques~\cite{Piazza2014Quantum}. 

\par

In this work we investigate quantum-statistical effects on the nonequilibrium stationary state close to the self-organisation threshold. Upon including quantum fluctuations and Markovian noise induced by photon losses, we compute the density matrix of the joint state of light and up to five atoms. Here the particles' average energy is not set by a prescribed temperature, instead their momentum distribution follows dynamically from the interplay of cavity cooling and fluctuations. In general, a nonthermal distribution emerges, which exhibits strong atom-field correlations. In principle, even starting from a particle ensemble at finite temperature, the combined cooling and self-ordering can lead to a quantum degenerate ordered final state.

\par

Since the photons leaking out of the cavity provide a directly accessible non-destructive probe of the system, we characterise the self-organisation transition with the help of the Husimi-$Q$-function, the intensity and the power spectrum of the light emitted from the cavity. This provides an experimentally directly accessible tool, so that the whole self-ordering phase transition process can be monitored in real time even in a single experimental run with minimal backaction~\cite{Mekhov2012Quantum}. To investigate the full quantum limit of cavity-induced self-organisation and cooling we resort to numerical solutions of the master equation within a truncated atomic momentum Hilbert space for a finite fixed particle number. 

\section{Effective mode model}\label{sec:model}

We consider $N$ ultracold particles within a high-$Q$ optical resonator illuminated by a transverse laser far detuned from any internal resonance (\cf fig.~\ref{fig:system}).
The light scattered by the particles into the resonator interferes with the pump field and creates an optical lattice potential along the cavity axis. We assume the particles' motion to be restricted to the cavity axis by means of a suitable transversal confinement. The coherent time evolution of this joint particle-field system is then governed by the second-quantised Hamiltonian~\cite{Konya2011,Ritsch2013}
\begin{multline}
  H=\int_{-L/2}^{L/2}\hat{\Psi}^\dagger(x)\Bigl[ -\frac{\hbar^2 }{2m} \frac{\mathrm{d}^2}{\mathrm{d}x^2}+\hbar U_0 a^\dagger a \cos ^2(kx)\Bigr.\\
  \Bigl.+\hbar \eta (a+a^\dagger)\cos (kx) \Bigr]\hat{\Psi}(x)\,\mathrm{d}x
  -\hbar \DC a^\dagger a,
  \label{eq_Hsq}
\end{multline}
where $k$ denotes the wave number of the cavity mode which also sets the recoil energy scale $\hbar\omrec=\hbar^2 k^2/(2m)$,  $m$ is the mass of a single particle, and $L=M2\pi/k$ is the unit cell length of the periodic boundary conditions, equal to an integer
$M$ times the wavelength $\lambda=2\pi/k$. The parameter $U_0$ is the light shift per particle, $\eta$ the effective pump amplitude per atom and $a$ the annihilation operator of a cavity photon.
The field operators $\hat{\Psi}^\dagger(x)$ ($\hat{\Psi}(x)$) create (destroy) a particle at position $x$ and obey bosonic or fermionic commutation or anticommutation relations, respectively.
The Hamiltonian~\eqref{eq_Hsq} is expressed in a reference frame rotating with the pump laser frequency $\omega_\mathrm{p}$. Consequently, the detuning $\DC:=\omega_\mathrm{p}-\omega_\mathrm{c}$ between the pump and the bare cavity resonance frequency explicitly appears in eq.~\eqref{eq_Hsq}. Note that this effective model describes any kind of small enough linearly polarisable particles~\cite{Ritsch2013}.

\par

As an important feature the Hamiltonian~\eqref{eq_Hsq} allows for momentum transfers of $\Delta p=\pm\hbar k$ in addition to $\Delta p=\pm2\hbar k$ from scattering within the cavity~\cite{Niedenzu2012,Sandner2013subrecoil}. This enables the system to self-order and has also important consequences for the cooling behaviour.
Note that direct particle-particle interactions, which could induce arbitrary momentum transfers $\Delta p$ (\eg collisions), are neglected.

\par

The second-quantised representation~\eqref{eq_Hsq} allows us to optimally exploit the symmetries of the system and the bosonic or fermionic statistics of the quantum particles. To this end we expand the particle's creation and annihilation operators in the complete set of orthonormal mode functions
\begin{equation}\label{eq_modefunctions}
  \left\{\sqrt{\frac{1}{L}},\sqrt{\frac{2}{L}}\cos (n\deltak x),\sqrt{\frac{2}{L}}\sin (n\deltak x)\right\}
\end{equation}
with integer $n\ge 1$. The momentum space discretisation is given by $\hbar\deltak=\hbar2\pi/L$, therefore the cavity wave vector $k$ is an integer multiple of $\deltak$ ($k=M\deltak$).
As discussed in~\cite{Konya2011} the high-energy modes will be excluded from the model by introducing a cutoff $n\le \nmax$. Note that the subspaces spanned by odd (sine) and even (cosine) parity modes, respectively, are decoupled. For bosons initially in a BEC the sine modes remain unpopulated and can be excluded from the picture.  For a degenerate Fermi gas, however, all modes with energies below the Fermi energy are filled up, therefore sine and cosine modes both have to be taken into account. Nevertheless, the dimensionality of the actual problem to be solved can be further reduced by exploiting invariant subspaces (\cf appendix~\ref{sec:modemodel}). Additionally, we introduce a cutoff for the cavity photon number.

\par

Photons leaking out of the resonator inevitably introduce noise and damping into the system and thus prevent a pure Hamiltonian evolution of the system's Schr\"odinger wave function. Much rather, the system will evolve into a stochastic mixture described by the joint particle-field density operator governed by the master equation~\cite{Gardiner2000}
\begin{equation}\label{eq_master}
  \dot\rho = \frac{1}{i\hbar}\left[H,\rho\right]+\kappa\left(2a\rho a^\dagger-a^\dagger a\rho-\rho a^\dagger a\right),
\end{equation}
with the cavity photon decay rate $2\kappa$. We neglect spontaneous emission of the atoms under the assumption of a large detuning of the laser with respect to any internal atomic resonance. 

\par 
An expansion of the (bosonic or fermionic) field operator $\hat{\Psi}(x)$ in terms of the mode functions~\eqref{eq_modefunctions} yields an effective mode model, whose explicit Hamiltonian is reported in appendix~\ref{sec:modemodel}. 
We restrict ourselves to only the appreciably occupied modes, to be able to integrate the master equation~\eqref{eq_master} directly. This is possible since the occupation of the higher-energy modes remains low due to the ongoing cavity cooling effect~\cite{domokos2002collective}. 

\section{Self-organisation in the quantum regime}
\label{sec:self-organization}

Self-organisation can be understood already on a classical level, \ie with classical particles and a damped cavity field~\cite{domokos2002collective,Griesser2010}.
At this level, when the threshold pump strength for the phase transition is reached, the symmetry of the system is spontaneously broken as the initially homogeneously distributed particles organise and occupy one of two possible configurations~\cite{domokos2002collective,Ritsch2013} corresponding to even or odd sites of the cavity optical lattice $\sim\cos^2(kx)$. In the perfectly homogeneous phase, no photons are scattered into the cavity because of destructive interference and self-organisation is triggered only by density fluctuations. The occupation of the odd or even potential wells, respectively, is associated with one of two opposite phases of the intracavity light field relative to the pump laser. The order parameter of the phase transition is given by $\Theta = \langle \cos (kx) \rangle $ (or equivalently $\langle a\rangle$), which is zero in the completely homogeneous phase and $\pm1$ in the limit that the particles are completely localised at antinodes.
Quantum-mechanically, the dynamics lead towards a self-organised state which is a superposition of the two configurations in odd and even wells correlated with the corresponding field phase~\cite{Maschler2007Entanglement}.  Initial atom-field entanglement decays on the timescale of the photon lifetime rendering the stationary state into a mixture of the two configurations.

\par

For both the entangled state and the mixture the mean field $\langle a\rangle$ and the order parameter $\langle \cos(kx) \rangle$ are zero, while the photon number $\langle a^\dagger a\rangle$ is finite. Above threshold, the stationary state is approximately~\cite{Vukics2007a}
\begin{align}
  \rhostationary\approx\frac{1}{2} \ket{\alpha,+}\bra{\alpha,+}+\frac{1}{2}\ket{-\alpha,-}\bra{-\alpha,-} ,
  \label{eq:stationarystate}
\end{align}
where $\ket{\alpha}$ is a coherent state with complex field amplitude $\alpha$ and $\ket{\pm}$ is the state of particles organised in the odd or even wells, respectively. With increasing pump strength, the components' coherent states $\ket{\alpha}$ and $\ket{-\alpha}$ have less and less overlap.
\begin{figure}%
  \centering%
  \includegraphics{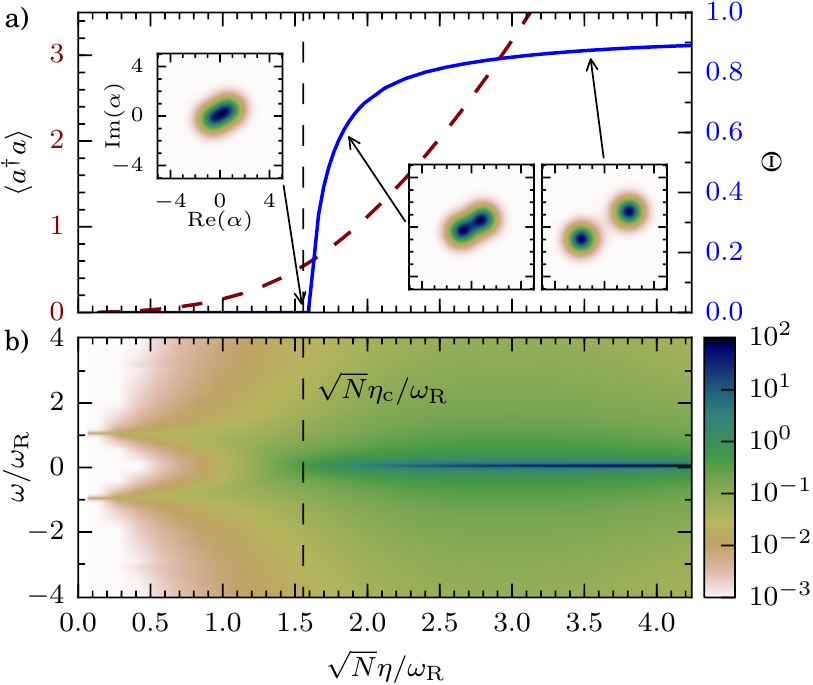}%
  \caption{a) The order parameter $\Theta$ and intracavity photon number $\braket{a^\dagger a}$, b) the cavity field spectrum $|S(\omega)|$ as a function of the pump strength, respectively. Insets: The $Q$-function of the photon field splits up into two coherent peaks across the threshold. Parameters: $N=2$ bosons, $U_0/\omrec=0$, $\kappa/\omrec=1$, $\DC/\omrec=-1.75$.}%
  \label{fig:order_spectrum}%
\end{figure}%
This behaviour can be seen in the $Q$-function~\cite{wallsbook} of the cavity field, as shown in fig.~\ref{fig:order_spectrum}a.

\par

The classical order parameter $\langle \cos(kx)\rangle_{\rhostationary}$ is always zero and thus unsuitable in the quantum case.
In order to find an alternative measure, we project $\rhostationary$ to obtain
$\rhoalpha\coloneq \braket{\alpha|\rhostationary|\alpha}/\tr \braket{\alpha|\rhostationary|\alpha}$, with $\alpha$ maximising the $Q$-function obtained from the cavity field associated with $\rhostationary$.
Far above threshold, where $\braket{\alpha|-\alpha}\approx 0$, we have $\rho_\alpha\approx \ket{+}\bra{+}$. We then calculate the order parameter with this projected particle state, \ie 
\begin{equation}
  \Theta:=\langle \cos(kx)\rangle_{\rhoalpha}.
  \label{eq:theta}
\end{equation}
As long as the $Q$-function only has one maximum at $\alpha=0$, this is exactly zero because of the system's symmetry. The threshold $\etac$, \ie the driving for which $\Theta$ first becomes larger than zero, is therefore \emph{defined} to be the pump strength at which the $Q$-function splits up and has two local maxima. The sudden increase of the mean intracavity photon number at the threshold, which is expected in the thermodynamic limit, is smoothened in the few particle regime considered here.
This is illustrated in fig.~\ref{fig:order_spectrum}a.

\par

In addition to the order parameter, the onset of self-organisation can also be seen in the cavity field spectrum, as observed experimentally with a BEC~\cite{Mottl2012Roton,Landig2015Measuring}. To this end we calculate $S(\omega)=\mathcal{F}(\langle a^\dagger(T+t)a(T)\rangle)$, where $\mathcal{F}$ is the Fourier transform with respect to $t$, and $T$ is a time sufficiently large for the system to have reached its stationary state~\cite{wallsbook}. Figure~\ref{fig:order_spectrum}b shows $\abs{S(\omega)}$ for bosons. Below threshold, the two peaks correspond to collective modes, characterised by light-field oscillations and particles excited from momentum zero to $\hbar k$ and vice versa. At low pump strength, the peaks are located at $\sim\pm \omrec$ and are almost perfectly sharp (undamped) since the collective modes are almost purely atomic excitations.  With increasing pump strength, apart from a broadening of the peaks due to mixing of atomic excitations with the decaying photonic excitations, we observe a softening of the collective modes which become energetically less and less costly. At a critical pump strength these modes are shifted to zero energy ($\omega=0$ in fig.~\ref{fig:order_spectrum}b). This process, accompanied by collective scattering of photons into the cavity at the laser frequency, is called superradiant because of the quadratic scaling of the photon number with $N$. The presence of such a dominant coherent peak at $\omega=0$ is thus an additional characterisation of self-organisation, as already observed semi-classically~\cite{Schuetz2014Prethermalization}. Within our model this superradiant coherent peak has zero width since we do not include a finite linewidth of the pump laser.

\par

For fermions, the behaviour of the order parameter is qualitatively similar to the bosonic case illustrated in fig.~\ref{fig:order_spectrum}a.  The spectrum, however, shows qualitative differences which will be discussed below.

\section{Quantum statistics and self-organisation}

Self-organisation and superradiance have been demonstrated with both a thermal gas~\cite{Black2003observation,Arnold2012self} in the classical regime as well as for a BEC  close to $T=0$~\cite{Baumann2010dicke,Kessler2014Steering}.  
\begin{figure}%
  \centering%
  \includegraphics{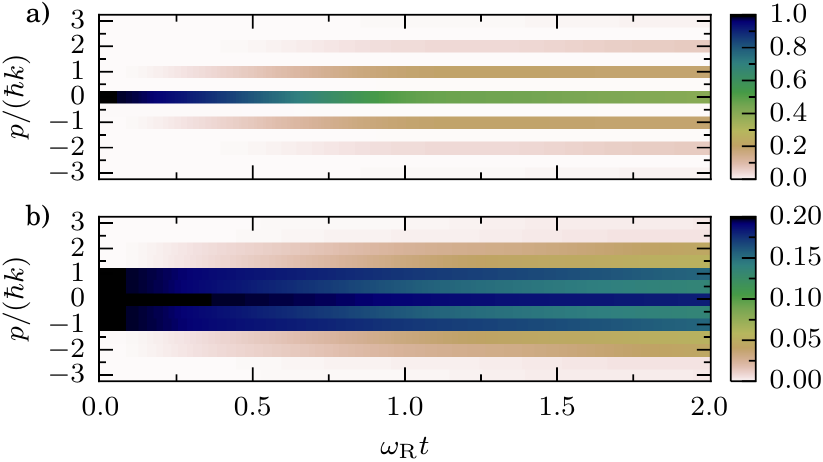}%
  \caption{Time evolution of the momentum population $p$ for a) bosons and b) fermions, starting with a quantum degenerate initial state at temperature zero. Participation of the fermion with $p=0$ in the photon scattering dynamics is strongly suppressed, as its target states $p/(\hbar k)=\pm 1$ are essentially blocked.  Parameters: $N=5$, $\eta/\omrec=2$, $\DC/\omrec=-4$, $U_0/\omrec=-0.5$, $\deltak/k=0.5$. }%
  \label{fig:dynamics}%
\end{figure}%
\begin{figure}[tbp]%
  \centering%
  \includegraphics{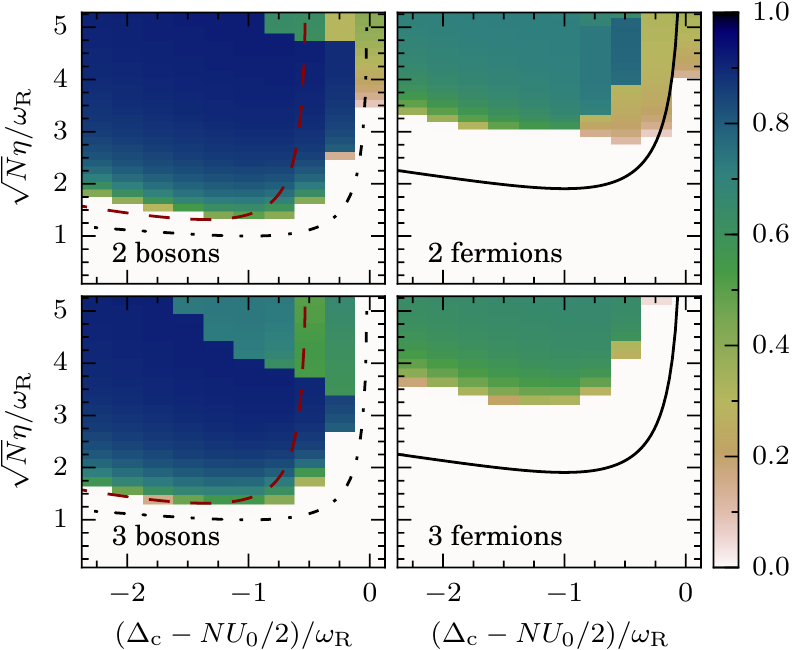}%
  \caption{The phase diagram of two and three bosons and fermions, respectively. The threshold is compared with the analytic results of~\cite{Niedenzu2011kinetic} in the semiclassical limit (dashed) and~\cite{Nagy2008self} for bosons (dash-dotted), as well as~\cite{Piazza2014Quantum} for fermions (solid, $\kfermi/k=1$). For bosons, the sudden decrease of the order parameter with increasing pump strength is an artefact of the photon number cutoff. Parameters: $U_0/\omrec=-0.5$, $\kappa/\omrec=1$.}%
  \label{fig:phasediagram}%
\end{figure}%
\begin{figure*}[tbp]%
  \centering%
  \includegraphics{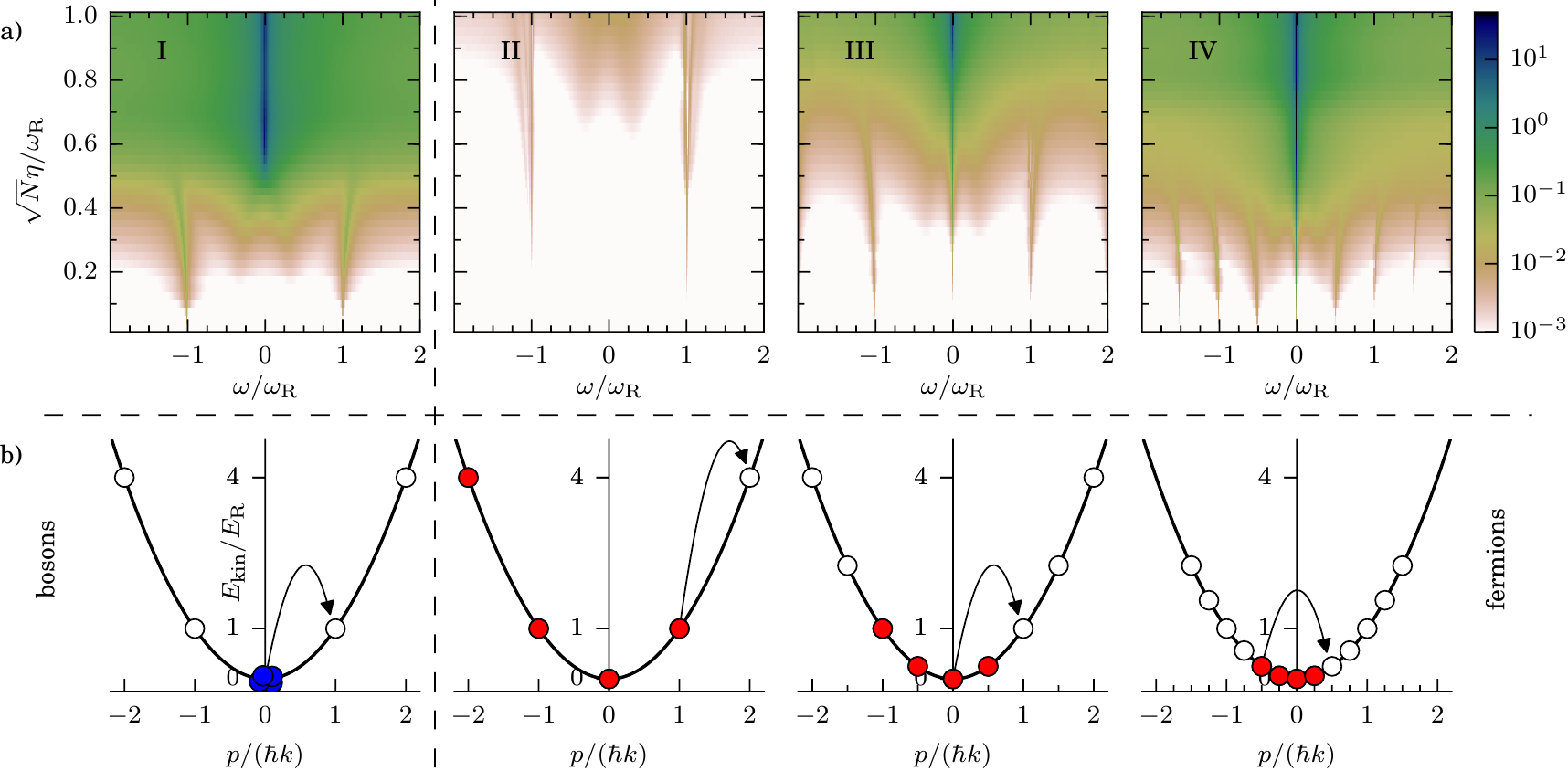}%
  \caption{a) Cavity field spectrum and b) initial conditions with lowest-energy particle excitations for bosons (I) and fermions (II--IV). For fermions the value $\deltak/k=1,0.5,0.25$ is varied. For bosons only the case $\deltak/k=1$ is shown, as the dynamics are unchanged in the other cases. Note that in case II the peaks in the spectrum at $\pm \omrec$ correspond to transitions which are forbidden in the initial state but become possible during relaxation towards the stationary state. The frequency of the transition indicated by the arrow in II b) lies outside the depicted spectral range. The resonance condition $k=2\kfermi$ is fulfilled in IV.  Here the central peak in the spectrum at the laser frequency ($\omega=0$) is dominant even for the smallest pump strengths. Parameters: $N=4$, $U_0/\omrec=-1/16$, $\kappa/\omrec=1/8$, $\DC/\omrec=-1/2$.}%
  \label{fig:umklapp}%
\end{figure*}%
Self-ordering of course requires excitations out of the $p=0$ condensed state to higher momenta with increments $\pm\hbar k$ (\cf fig.~\ref{fig:dynamics}a).
The qualitative behaviour of BEC self-organisation resembles its thermal counterpart, with the recoil energy scale $\hbar\omrec$ playing the role of the temperature scale $\kB T$~\cite{Nagy2008self}.

\par

In contrast to bosons, fermionic statistics allows for a qualitatively different self-ordering behaviour~\cite{Keeling2014Fermionic,Piazza2014Umklapp,Chen2014Superradiance}.
In fact, a Fermi gas introduces a new internal energy scale, the Fermi energy $\efermi$, and indeed its ratio  to the recoil energy $\hbar\omrec$ dramatically affects the behaviour of the fermionic self-organisation. 
As  illustrated in fig.~\ref{fig:dynamics}b, due to the Pauli pressure fermions occupy higher momentum states even at $T=0$ including cosine and sine modes.
This prevents a direct mapping of the system's Hamiltonian onto an effective Dicke model (demonstrated for a BEC~\cite{Nagy2010Dicke}) and the understanding of self-organisation in terms of a simple superradiant transition is not possible, except for the limiting case of an infinitely large Fermi momentum relative to the cavity photon momentum~\cite{Zubairy2014Dicke}.

\par

The central effects of Fermi statistics on self-ordering so far have been evaluated based on the assumption of a thermal equilibrium for the particles and neglecting the role of quantum fluctuations of the cavity field~\cite{Keeling2014Fermionic,Piazza2014Umklapp,Chen2014Superradiance}. Here we test these predictions in comprehensive nonequilibrium simulations for a small number of particles including light-matter entanglement and quantum fluctuations of the cavity field.
Due to Pauli blocking we can expect that for fermions not all particles will equally participate in the self-organisation phenomena. Whilst for classical particles and bosons scattering of a photon is always allowed and thus the critical pump strength scales as $\eta_{\rm c}\propto 1/\sqrt{N}$~\cite{Asboth2005selforganization,Nagy2008self,Niedenzu2011kinetic,Piazza2013bose}, only fermions coupled to an unoccupied final state can scatter (\cf fig.~\ref{fig:dynamics}b). This fraction strongly depends on the Fermi energy and hence scales differently with $N$. For a sharp Fermi surface, $\etac$ can be estimated analytically. In one spatial dimension one obtains $\epsilon_{\rm F}\propto (N/L)^2$, which in turn means $\eta_{\rm c}\propto \sqrt{\ln(N)/N}$ for $k\gg  k_{\rm F}$ or even $\eta_{\rm c}\propto \sqrt{N}$ for $k \ll  k_{\rm F}$~\cite{Piazza2014Umklapp}. 

\par

Figure~\ref{fig:phasediagram} depicts the value of the order parameter~\eqref{eq:theta} as a function of the (dispersively shifted) cavity detuning and the pump amplitude for two and three bosons and fermions, respectively. In this low-particle regime the notion of a sharp Fermi surface becomes questionable. Nevertheless, we see qualitative agreement of our numerical data with the scaling discussed above: for $N=3$ fermions the rescaled $\sqrt{N}\etac$ is larger than for $N=2$, while for bosons it does not change.

\par

While the Fermi blockade causes an unfavourable threshold scaling with the number of particles, the existence of a Fermi surface can also give rise to a resonant reduction of the threshold if $k=2k_{\rm F}$.
This ``umklapp superradiance''~\cite{Piazza2014Umklapp} involves particles scattered from one end of the Fermi surface to the other (\ie from $-\kfermi$ to $\kfermi$) by the two-photon transition with momentum transfer $\hbar k=2\hbar \kfermi$. This process has essentially no energy cost and thus can drastically lower the self-organisation threshold.  Here the ordered state is very close to the Fermi surface and has essentially the same kinetic energy as the homogeneous ground state. Note that this availability of some already excited particles can also explain the threshold reduction found in~\cite{Piazza2013bose} for bosons at small temperature compared to the zero temperature limit.

\par

In our simulations with a fixed number of fermions prepared initially at $T=0$, the resonance condition $k=2\kfermi$ can be reached by different means: either by adjusting the cavity wavenumber $k$ directly or by altering the particle density $L/N$ and thus the Fermi momentum. Here we choose to keep $k$ fixed and vary the trap volume $L$ which the fermions can occupy by altering $\deltak=2\pi/L$. This has the advantage that the recoil frequency $\omrec$ stays the same for all parameters.

\par

In order to investigate the role of umklapp superradiance, we prepare the initial state in the resonant condition, perform the full quantum dissipative evolution and compute the order parameter and intracavity spectrum once the stationary state is reached. As mentioned in the beginning, our simulations include the relaxation of the particles into a (generally) nonequilibrium stationary state different from the initial one. In particular, redistribution of energy and momentum will alter the Fermi surface, thereby modifying the resonance condition. Whether and how the umklapp superradiance manifests itself in this stationary state is a non-trivial question. 
The answer is illustrated in fig.~\ref{fig:umklapp}.
Here we compare bosons in a BEC with zero-temperature fermions at three different densities corresponding to the three Fermi momenta $\kfermi=2k$ ($\deltak=k$), $\kfermi=k$ ($\deltak=0.5k$) and $2\kfermi=k$ ($\deltak=0.25k$). In the last case the resonance condition is fulfilled. Initially, with no cavity photons present, the particles move freely with a quadratic dispersion relation. As already mentioned, the cavity only supports momentum transfer to and from the particles in units $\hbar k$. This corresponds to steps of $1$ on the $x$--axes of fig.~\ref{fig:umklapp}b, where arrows indicate the particle excitations with lowest possible energy cost. These excitations are visible in the spectra shown in~\ref{fig:umklapp}a as peaks at the corresponding frequencies. 

\par

The effect of resonant umklapp scattering is clearly visible in the spectra by comparing the initially resonant case (fig.~\ref{fig:umklapp}a IV) with the other two fermionic non-resonant cases as well as with the bosonic case. The appearance of a coherent peak at zero frequency in the resonant case signals the onset of superradiance even at almost vanishing pump strengths.  For all the other cases this onset of a coherent scattering peak is located at higher pump strengths.
Such a low superradiant threshold is possible because of (almost) resonant particle excitations with momentum $\hbar k$, creating the correct density modulation period for coherent Bragg scattering into the cavity.

\par

Even though the umklapp resonant condition can lower the threshold for the appearance of the superradiant peak much below the bosonic threshold, the number of photons in the cavity and the amplitude of the order parameter~\eqref{eq:theta} can still be lower than in the bosonic case, due to the unfavourable scaling with $N$ illustrated above. For instance, the threshold pump strength scales as $\eta_{\rm c}\sim \sqrt{N/\ln(N)}$ at $k\simeq 2k_{\rm F}$ if we assume a sharp Fermi surface. Therefore the effect of umklapp superradiance is much less visible in the photon number or the order parameter as compared to the spectrum. 
\begin{figure}%
  \centering%
  \includegraphics{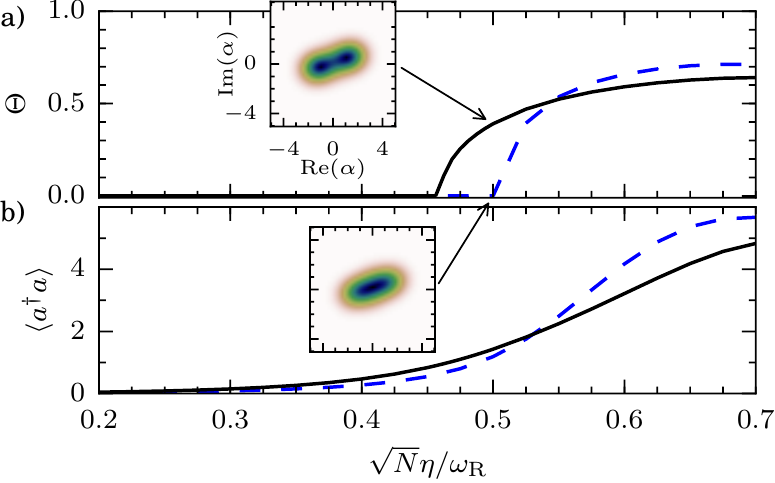}%
  \caption{a) Order parameter and b) photon number for fermions (solid) and bosons (dashed) for the resonance case $k=2\kfermi$. The threshold for fermions is lower compared to bosons. Insets: The $Q$-function of the photon field at a pump strength which lies above threshold for fermions and below threshold for bosons, respectively. Same parameters as in fig.~\ref{fig:umklapp}.}%
  \label{fig:orderparameter_resonance}%
\end{figure}%
Still, for certain parameter regimes and in the resonant case a lower self-organisation threshold for fermions can be seen even in these quantities, as shown in fig.~\ref{fig:orderparameter_resonance}.  

\section{Conclusions}

Self-ordering of ultracold particles in an optical resonator field survives as a stationary nonthermal equilibrium state in the long-time limit even if cavity losses with their inherent field quantum fluctuations are taken into account. The quantum statistics of the particles turns out to  have a decisive influence on the light-induced crystallisation of quantum particles in cavity fields even in this nonequilibrium context. This can be nicely monitored in real time with minimal perturbations by observing the spectrum of the scattered light.  

\begin{acknowledgments}
We acknowledge support by the EU ITN project CCQED, the Austrian Science Fund FWF through project SFB FoQuS F4013 and the Austrian Ministry of Science BMWF as part of the UniInfrastrukturprogramm of the Focal Point Scientific Computing at the University of Innsbruck. FP acknowledges support from the APART Fellowship of the Austrian Academy of Sciences.
\end{acknowledgments}

\begin{appendix}
\section{Mode model}\label{sec:modemodel}
Starting from the Hamiltonian~\eqref{eq_Hsq},
we use the mode model from~\cite{Konya2011}, extended to include the sine modes and generalized for an arbitrary momentum space discretisation $\deltak$.

We expand the particles' field operator $\hat{\Psi}(x)$ in the complete set of orthonormal mode functions
\begin{align}\label{eq_modefunctions_appendix}
  \varphi_{0,1}(x)&=\sqrt{\frac{1}{L}}\notag\\
  \varphi_{n,1}(x)&=\sqrt{\frac{2}{L}}\cos (n\deltak x), & \varphi_{n,-1}(x)&=\sqrt{\frac{2}{L}}\sin (n\deltak x),
\end{align}
with integer $n\ge 1$. The modes are labeled by the multi-index $\boldsymbol{j}=(j,j_{\pi})$, which contains the mode number $j$ and the parity $j_{\pi}=1$ for the cosine and $j_{\pi}=-1$ for the sine modes. This yields the decomposition
\begin{align}
  \hat{\Psi}(x)&=\smashoperator{\sum_{\boldsymbol{j},j\le n_c}}\varphi_{\boldsymbol{j}}(x) \hat{c}_{\boldsymbol{j}},
\end{align}
where the $\hat{c}_{\boldsymbol{j}}$ follow either bosonic or fermionic (anti-) commutation rules and the $\hat{c}_{j,-1}$ operators are formally set to zero for bosons.

\par

With $\hat{\boldsymbol{c}}=(\hat{c}_{0,1},\dots,\hat{c}_{\nmax,1},\hat{c}_{1,-1},\dots,\hat{c}_{\nmax,-1})^T$ as column vector and $\hat{\boldsymbol{c}}^\dagger=(\hat{c}_{0,1}^\dagger,\dots,\hat{c}_{\nmax,1}^\dagger,\hat{c}_{1,-1}^\dagger,\dots,\hat{c}_{\nmax,-1}^\dagger)$ as row vector, we can write the Hamiltonian~\eqref{eq_Hsq} as
\begin{multline}
  H=-\hbar\DC a^\dagger a +\hbar\omrec \left( \hat{\boldsymbol{c}}^\dagger \boldsymbol{K} \hat{\boldsymbol{c}} \right)
  +\hbar \eta(a+a^\dagger)\left(\hat{\boldsymbol{c}}^\dagger \boldsymbol{C}_{k/\deltak} \hat{\boldsymbol{c}}\right)\\
+\hbar \frac{U_0}{2} a^\dagger a\left(\hat{\boldsymbol{c}}^\dagger \left(\boldsymbol{I}+\boldsymbol{C}_{2k/\deltak} \right)\hat{\boldsymbol{c}}\right),
  \label{eq_Hsqmode}
\end{multline}
where $\omrec=\hbar k^2/(2m) $ is the recoil frequency and $\boldsymbol{I}$ is the identity matrix. The matrices $\boldsymbol{K}$ and $\boldsymbol{C}_{M}$ with integer $M$, producing the kinetic energy term and the terms proportional to $\cos (kx)$ and $\cos (2kx)$, respectively, are defined as
\begin{equation}
    \boldsymbol{K}=\frac{\deltak^2}{k^2}
\left(
\def\objectstyle{\scriptstyle}
\vcenter{\xymatrix@=1pt@M=0pt{
0\\
&1^2\\
&&2^2 \ar@{.}[ddrr]\\
\\
&&&&\nmax^2\\
&&&&&1^2 \ar@{.}[ddrr]\\
\\
&&&&&&&\nmax^2
}
}
\right),\ 
\boldsymbol{C}_M=
\left(
\def\objectstyle{\scriptstyle}
\def\labelstyle{\scriptscriptstyle}
\vcenter{\xymatrix@=5px@M=0pt@!0{
&&&&a\\
&&&b&&b\\
&&&&&&b\ar@{.}[dr] \\
&b\ar@{.}[uurr]& &&&&&\\
a&&&&\\
&b&&&\\
&&b\ar@{.}[dr]\ar@{<->}|{M}[uuuurrrr]\\
&&&&&&&\\
&&&&&&&&&&-b&&b\\
&&&&&&&&&&&&&b\\
&&&&&&&&-b\ar@{.}[uurr]&&&&&&b\ar@{.}[dr] \\
&&&&&&&&&& &&&&&\\
&&&&&&&&b&&&&\\
&&&&&&&&&b&&&\\
&&&&&&&&&&b\ar@{.}[dr]\ar@{<->}|{M}[uuuurrrr]\\
&&&&&&&&&&&&&&&
\save "8,8"."1,1"*+<3px>[F.]\frm{} \restore
\save "16,16"."9,9"*+<3px>[F.]\frm{}\restore
}
}
\right)
\label{eq_matrices}
,
\end{equation}
where $a=1/\sqrt{2}$ and $b=1/2$. For $\boldsymbol{C}_M$, the two blocks belong to the cosine and sine modes, respectively. The two populated secondary diagonals are separated by $M$ modes. Additionally, modes $j,k$ for which $j+k=M$ are coupled by the entries $b$ for cosine modes and $-b$ for sine modes, respectively. For reference, the quadratic forms appearing in the mode-expanded Hamiltonian~\eqref{eq_Hsqmode} read
\begin{multline}
  \hat{\boldsymbol{c}}^\dagger \boldsymbol{C}_M \hat{\boldsymbol{c}}= \frac{1}{\sqrt{2}} \left( \hat{c}_{M,1}^{\dagger} \hat{c}_{0,1}+ \hat{c}_{0,1}^\dagger c_{M,1}\right)\\+\frac{1}{2}\ \smashoperator{\sum_{\substack{\boldsymbol{j},\boldsymbol{k}\\1\le j,k\le \nmax}}}
  \delta_{j_{\pi}k_{\pi}}
  \bigl[ \delta_{j,k+M}
  +\delta_{j+M,k}+j_\pi\delta_{j+k,M} \bigr] \hat{c}_{\boldsymbol{j}}^\dagger \hat{c}_{\boldsymbol{k}}.
\end{multline}
For our simulations we use the usual basis of Fock states
\begin{equation}
  \Bigl\{ \ket{n_{0,1},\dots,n_{\nmax,1};n_{1,-1},\dots,n_{\nmax,-1}} \Bigr\}_{\sum_{\boldsymbol{j}}n_{\boldsymbol{j}}=N},
  \label{eq_fock_basis}
\end{equation}
where $n_{\boldsymbol{j}}=0,\dots,N$ for bosons and $n_{\boldsymbol{j}}=0,1$ for fermions.

As already noted and as can be seen from the coupling matrices~\eqref{eq_matrices}, sine and cosine modes are decoupled. However, for $M\coloneq k/\deltak>1$ one can identify further subgroups of modes which are decoupled from all other groups. We will now specify these groups and use them to identify invariant subspaces of the Hilbert space $\mathcal{H}$ spanned by the basis states~\eqref{eq_fock_basis}.

Let us group all available modes into $s$ sets of modes, where $s\coloneq \floor{M/2}+1$ for bosons and $s\coloneq 2\floor{M/2}+1$ for fermions:
\begin{equation}
  g_{\boldsymbol{l}}\coloneq \left\{ \boldsymbol{j}\big|j_\pi=l_\pi,j\bmod M=l\vee j\bmod M=M-l\right\},
  \label{eq_groups}
\end{equation}
with $l_\pi=1$ ($l_\pi=\pm 1$) for bosons (fermions) and $l\leq\floor{M/2}$. The significance of these sets is that two modes are coupled through the dynamics induced by the Hamiltonian~\eqref{eq_Hsqmode} if and only if they belong to the same set. As a consequence, the number operator corresponding to each group, \ie
\begin{equation}
  \hat{n}_{\boldsymbol{l}}\coloneq\sum_{\boldsymbol{j}\in g_{\boldsymbol{l}}}
  \hat{c}_{\boldsymbol{j}}^\dagger \hat{c}_{\boldsymbol{j}}
  \label{eq_number_group}
\end{equation}
commutes with $H$ and $\expect{\hat{n}_{\boldsymbol{l}}}$ is constant over time. This enables us to define invariant subspaces of $\mathcal{H}$: two basis states of~\eqref{eq_fock_basis} belong to the same invariant subspace if and only if they produce the same $s$-tuple $(\expect{\hat{n}_{\boldsymbol{l}}})_{\boldsymbol{l}}$. We can exclude any basis state from our model if it belongs to a subspace which is not populated by the initial state.

For example, for a degenerate Fermi gas with $N=4$, $k/\deltak=4$ and $\nmax=6$, the basis~\eqref{eq_fock_basis} consists of $715$ states. By virtue of the above considerations we only have to take $72$ basis states into account.

  
%
\end{appendix}

%


\begin{thebibliography}{45}%
\makeatletter
\providecommand \@ifxundefined [1]{%
 \@ifx{#1\undefined}
}%
\providecommand \@ifnum [1]{%
 \ifnum #1\expandafter \@firstoftwo
 \else \expandafter \@secondoftwo
 \fi
}%
\providecommand \@ifx [1]{%
 \ifx #1\expandafter \@firstoftwo
 \else \expandafter \@secondoftwo
 \fi
}%
\providecommand \natexlab [1]{#1}%
\providecommand \enquote  [1]{``#1''}%
\providecommand \bibnamefont  [1]{#1}%
\providecommand \bibfnamefont [1]{#1}%
\providecommand \citenamefont [1]{#1}%
\providecommand \href@noop [0]{\@secondoftwo}%
\providecommand \href [0]{\begingroup \@sanitize@url \@href}%
\providecommand \@href[1]{\@@startlink{#1}\@@href}%
\providecommand \@@href[1]{\endgroup#1\@@endlink}%
\providecommand \@sanitize@url [0]{\catcode `\\12\catcode `\$12\catcode
  `\&12\catcode `\#12\catcode `\^12\catcode `\_12\catcode `\%12\relax}%
\providecommand \@@startlink[1]{}%
\providecommand \@@endlink[0]{}%
\providecommand \url  [0]{\begingroup\@sanitize@url \@url }%
\providecommand \@url [1]{\endgroup\@href {#1}{\urlprefix }}%
\providecommand \urlprefix  [0]{URL }%
\providecommand \Eprint [0]{\href }%
\providecommand \doibase [0]{http://dx.doi.org/}%
\providecommand \selectlanguage [0]{\@gobble}%
\providecommand \bibinfo  [0]{\@secondoftwo}%
\providecommand \bibfield  [0]{\@secondoftwo}%
\providecommand \translation [1]{[#1]}%
\providecommand \BibitemOpen [0]{}%
\providecommand \bibitemStop [0]{}%
\providecommand \bibitemNoStop [0]{.\EOS\space}%
\providecommand \EOS [0]{\spacefactor3000\relax}%
\providecommand \BibitemShut  [1]{\csname bibitem#1\endcsname}%
\let\auto@bib@innerbib\@empty
\bibitem [{\citenamefont {Ritsch}\ \emph {et~al.}(2013)\citenamefont {Ritsch},
  \citenamefont {Domokos}, \citenamefont {Brennecke},\ and\ \citenamefont
  {Esslinger}}]{Ritsch2013}%
  \BibitemOpen
  \bibfield  {author} {\bibinfo {author} {\bibfnamefont {H.}~\bibnamefont
  {Ritsch}}, \bibinfo {author} {\bibfnamefont {P.}~\bibnamefont {Domokos}},
  \bibinfo {author} {\bibfnamefont {F.}~\bibnamefont {Brennecke}}, \ and\
  \bibinfo {author} {\bibfnamefont {T.}~\bibnamefont {Esslinger}},\ }\href
  {http://rmp.aps.org/abstract/RMP/v85/i2/p553_1} {\bibfield  {journal}
  {\bibinfo  {journal} {{Rev. Mod. Phys.}}\ }\textbf {\bibinfo {volume} {85}},\
  \bibinfo {pages} {553} (\bibinfo {year} {2013})}\BibitemShut {NoStop}%
\bibitem [{\citenamefont {Mekhov}\ and\ \citenamefont
  {Ritsch}(2012)}]{Mekhov2012Quantum}%
  \BibitemOpen
  \bibfield  {author} {\bibinfo {author} {\bibfnamefont {I.~B.}\ \bibnamefont
  {Mekhov}}\ and\ \bibinfo {author} {\bibfnamefont {H.}~\bibnamefont
  {Ritsch}},\ }\href {http://iopscience.iop.org/0953-4075/45/10/102001}
  {\bibfield  {journal} {\bibinfo  {journal} {{J. Phys. B}}\ }\textbf {\bibinfo
  {volume} {45}},\ \bibinfo {pages} {102001} (\bibinfo {year}
  {2012})}\BibitemShut {NoStop}%
\bibitem [{\citenamefont {Stamper-Kurn}(2014)}]{stamper2014cavity}%
  \BibitemOpen
  \bibfield  {author} {\bibinfo {author} {\bibfnamefont {D.~M.}\ \bibnamefont
  {Stamper-Kurn}},\ }in\ \href@noop {} {\emph {\bibinfo {booktitle} {{Cavity
  Optomechanics}}}}\ (\bibinfo  {publisher} {Springer},\ \bibinfo {year}
  {2014})\ pp.\ \bibinfo {pages} {283--325}\BibitemShut {NoStop}%
\bibitem [{\citenamefont {Horak}\ \emph {et~al.}(1997)\citenamefont {Horak},
  \citenamefont {Hechenblaikner}, \citenamefont {Gheri}, \citenamefont
  {Stecher},\ and\ \citenamefont {Ritsch}}]{Horak1997}%
  \BibitemOpen
  \bibfield  {author} {\bibinfo {author} {\bibfnamefont {P.}~\bibnamefont
  {Horak}}, \bibinfo {author} {\bibfnamefont {G.}~\bibnamefont
  {Hechenblaikner}}, \bibinfo {author} {\bibfnamefont {K.~M.}\ \bibnamefont
  {Gheri}}, \bibinfo {author} {\bibfnamefont {H.}~\bibnamefont {Stecher}}, \
  and\ \bibinfo {author} {\bibfnamefont {H.}~\bibnamefont {Ritsch}},\ }\href
  {http://prl.aps.org/abstract/PRL/v79/i25/p4974_1} {\bibfield  {journal}
  {\bibinfo  {journal} {{Phys. Rev. Lett.}}\ }\textbf {\bibinfo {volume}
  {79}},\ \bibinfo {pages} {4974} (\bibinfo {year} {1997})}\BibitemShut
  {NoStop}%
\bibitem [{\citenamefont {Aspelmeyer}\ \emph {et~al.}(2014)\citenamefont
  {Aspelmeyer}, \citenamefont {Kippenberg},\ and\ \citenamefont
  {Marquardt}}]{Aspelmeyer2014cavity}%
  \BibitemOpen
  \bibfield  {author} {\bibinfo {author} {\bibfnamefont {M.}~\bibnamefont
  {Aspelmeyer}}, \bibinfo {author} {\bibfnamefont {T.~J.}\ \bibnamefont
  {Kippenberg}}, \ and\ \bibinfo {author} {\bibfnamefont {F.}~\bibnamefont
  {Marquardt}},\ }\href {\doibase 10.1103/RevModPhys.86.1391} {\bibfield
  {journal} {\bibinfo  {journal} {{Rev. Mod. Phys.}}\ }\textbf {\bibinfo
  {volume} {86}},\ \bibinfo {pages} {1391} (\bibinfo {year}
  {2014})}\BibitemShut {NoStop}%
\bibitem [{\citenamefont {Wolke}\ \emph {et~al.}(2012)\citenamefont {Wolke},
  \citenamefont {Klinner}, \citenamefont {Ke{\ss}ler},\ and\ \citenamefont
  {Hemmerich}}]{Wolke2012}%
  \BibitemOpen
  \bibfield  {author} {\bibinfo {author} {\bibfnamefont {M.}~\bibnamefont
  {Wolke}}, \bibinfo {author} {\bibfnamefont {J.}~\bibnamefont {Klinner}},
  \bibinfo {author} {\bibfnamefont {H.}~\bibnamefont {Ke{\ss}ler}}, \ and\
  \bibinfo {author} {\bibfnamefont {A.}~\bibnamefont {Hemmerich}},\ }\href
  {http://www.sciencemag.org/content/337/6090/75.short} {\bibfield  {journal}
  {\bibinfo  {journal} {Science}\ }\textbf {\bibinfo {volume} {337}},\ \bibinfo
  {pages} {75} (\bibinfo {year} {2012})}\BibitemShut {NoStop}%
\bibitem [{\citenamefont {Sandner}\ \emph {et~al.}(2013)\citenamefont
  {Sandner}, \citenamefont {Niedenzu},\ and\ \citenamefont
  {Ritsch}}]{Sandner2013subrecoil}%
  \BibitemOpen
  \bibfield  {author} {\bibinfo {author} {\bibfnamefont {R.~M.}\ \bibnamefont
  {Sandner}}, \bibinfo {author} {\bibfnamefont {W.}~\bibnamefont {Niedenzu}}, \
  and\ \bibinfo {author} {\bibfnamefont {H.}~\bibnamefont {Ritsch}},\ }\href
  {http://stacks.iop.org/0295-5075/104/i=4/a=43001} {\bibfield  {journal}
  {\bibinfo  {journal} {{EPL (Europhys. Lett.)}}\ }\textbf {\bibinfo {volume}
  {104}},\ \bibinfo {pages} {43001} (\bibinfo {year} {2013})}\BibitemShut
  {NoStop}%
\bibitem [{\citenamefont {Elliott}\ \emph {et~al.}(2015)\citenamefont
  {Elliott}, \citenamefont {Mazzucchi}, \citenamefont {Kozlowski},
  \citenamefont {Caballero-Benitez},\ and\ \citenamefont
  {Mekhov}}]{Elliott2015Probing}%
  \BibitemOpen
  \bibfield  {author} {\bibinfo {author} {\bibfnamefont {T.~J.}\ \bibnamefont
  {Elliott}}, \bibinfo {author} {\bibfnamefont {G.}~\bibnamefont {Mazzucchi}},
  \bibinfo {author} {\bibfnamefont {W.}~\bibnamefont {Kozlowski}}, \bibinfo
  {author} {\bibfnamefont {S.}~\bibnamefont {Caballero-Benitez}}, \ and\
  \bibinfo {author} {\bibfnamefont {I.~B.}\ \bibnamefont {Mekhov}},\ }\href
  {http://arxiv.org/abs/1506.07700} {\bibfield  {journal} {\bibinfo  {journal}
  {arXiv preprint arXiv:1506.07700 [quant-ph]}\ } (\bibinfo {year}
  {2015})}\BibitemShut {NoStop}%
\bibitem [{\citenamefont {Domokos}\ and\ \citenamefont
  {Ritsch}(2002)}]{domokos2002collective}%
  \BibitemOpen
  \bibfield  {author} {\bibinfo {author} {\bibfnamefont {P.}~\bibnamefont
  {Domokos}}\ and\ \bibinfo {author} {\bibfnamefont {H.}~\bibnamefont
  {Ritsch}},\ }\href {\doibase 10.1103/PhysRevLett.89.253003} {\bibfield
  {journal} {\bibinfo  {journal} {{Phys. Rev. Lett.}}\ }\textbf {\bibinfo
  {volume} {89}},\ \bibinfo {pages} {253003} (\bibinfo {year}
  {2002})}\BibitemShut {NoStop}%
\bibitem [{\citenamefont {Asb\'{o}th}\ \emph {et~al.}(2005)\citenamefont
  {Asb\'{o}th}, \citenamefont {Domokos}, \citenamefont {Ritsch},\ and\
  \citenamefont {Vukics}}]{Asboth2005selforganization}%
  \BibitemOpen
  \bibfield  {author} {\bibinfo {author} {\bibfnamefont {J.~K.}\ \bibnamefont
  {Asb\'{o}th}}, \bibinfo {author} {\bibfnamefont {P.}~\bibnamefont {Domokos}},
  \bibinfo {author} {\bibfnamefont {H.}~\bibnamefont {Ritsch}}, \ and\ \bibinfo
  {author} {\bibfnamefont {A.}~\bibnamefont {Vukics}},\ }\href {\doibase
  10.1103/PhysRevA.72.053417} {\bibfield  {journal} {\bibinfo  {journal}
  {{Phys. Rev. A}}\ }\textbf {\bibinfo {volume} {72}},\ \bibinfo {pages}
  {053417} (\bibinfo {year} {2005})}\BibitemShut {NoStop}%
\bibitem [{\citenamefont {Grie{\ss}er}\ \emph {et~al.}(2010)\citenamefont
  {Grie{\ss}er}, \citenamefont {Ritsch}, \citenamefont {Hemmerling},\ and\
  \citenamefont {Robb}}]{Griesser2010}%
  \BibitemOpen
  \bibfield  {author} {\bibinfo {author} {\bibfnamefont {T.}~\bibnamefont
  {Grie{\ss}er}}, \bibinfo {author} {\bibfnamefont {H.}~\bibnamefont {Ritsch}},
  \bibinfo {author} {\bibfnamefont {M.}~\bibnamefont {Hemmerling}}, \ and\
  \bibinfo {author} {\bibfnamefont {G.~R.~M.}\ \bibnamefont {Robb}},\ }\href
  {http://link.springer.com/article/10.1140%2Fepjd%2Fe2010-00113-9?LI=true}
  {\bibfield  {journal} {\bibinfo  {journal} {{Eur. Phys. J. D}}\ }\textbf
  {\bibinfo {volume} {58}},\ \bibinfo {pages} {349} (\bibinfo {year}
  {2010})}\BibitemShut {NoStop}%
\bibitem [{\citenamefont {Niedenzu}\ \emph {et~al.}(2011)\citenamefont
  {Niedenzu}, \citenamefont {Grie{\ss}er},\ and\ \citenamefont
  {Ritsch}}]{Niedenzu2011kinetic}%
  \BibitemOpen
  \bibfield  {author} {\bibinfo {author} {\bibfnamefont {W.}~\bibnamefont
  {Niedenzu}}, \bibinfo {author} {\bibfnamefont {T.}~\bibnamefont
  {Grie{\ss}er}}, \ and\ \bibinfo {author} {\bibfnamefont {H.}~\bibnamefont
  {Ritsch}},\ }\href {http://stacks.iop.org/0295-5075/96/i=4/a=43001}
  {\bibfield  {journal} {\bibinfo  {journal} {{EPL (Europhys. Lett.)}}\
  }\textbf {\bibinfo {volume} {96}},\ \bibinfo {pages} {43001} (\bibinfo {year}
  {2011})}\BibitemShut {NoStop}%
\bibitem [{\citenamefont {Grie{\ss}er}\ \emph {et~al.}(2012)\citenamefont
  {Grie{\ss}er}, \citenamefont {Niedenzu},\ and\ \citenamefont
  {Ritsch}}]{Griesser2012cooperative}%
  \BibitemOpen
  \bibfield  {author} {\bibinfo {author} {\bibfnamefont {T.}~\bibnamefont
  {Grie{\ss}er}}, \bibinfo {author} {\bibfnamefont {W.}~\bibnamefont
  {Niedenzu}}, \ and\ \bibinfo {author} {\bibfnamefont {H.}~\bibnamefont
  {Ritsch}},\ }\href {http://stacks.iop.org/1367-2630/14/i=5/a=053031}
  {\bibfield  {journal} {\bibinfo  {journal} {{New J. Phys.}}\ }\textbf
  {\bibinfo {volume} {14}},\ \bibinfo {pages} {053031} (\bibinfo {year}
  {2012})}\BibitemShut {NoStop}%
\bibitem [{\citenamefont {Sch\"{u}tz}\ \emph {et~al.}(2013)\citenamefont
  {Sch\"{u}tz}, \citenamefont {Habibian},\ and\ \citenamefont
  {Morigi}}]{Schuetz2013cooling}%
  \BibitemOpen
  \bibfield  {author} {\bibinfo {author} {\bibfnamefont {S.}~\bibnamefont
  {Sch\"{u}tz}}, \bibinfo {author} {\bibfnamefont {H.}~\bibnamefont
  {Habibian}}, \ and\ \bibinfo {author} {\bibfnamefont {G.}~\bibnamefont
  {Morigi}},\ }\href {\doibase 10.1103/PhysRevA.88.033427} {\bibfield
  {journal} {\bibinfo  {journal} {{Phys. Rev. A}}\ }\textbf {\bibinfo {volume}
  {88}},\ \bibinfo {pages} {033427} (\bibinfo {year} {2013})}\BibitemShut
  {NoStop}%
\bibitem [{\citenamefont {Sch\"{u}tz}\ and\ \citenamefont
  {Morigi}(2014)}]{Schuetz2014Prethermalization}%
  \BibitemOpen
  \bibfield  {author} {\bibinfo {author} {\bibfnamefont {S.}~\bibnamefont
  {Sch\"{u}tz}}\ and\ \bibinfo {author} {\bibfnamefont {G.}~\bibnamefont
  {Morigi}},\ }\href {\doibase 10.1103/PhysRevLett.113.203002} {\bibfield
  {journal} {\bibinfo  {journal} {{Phys. Rev. Lett.}}\ }\textbf {\bibinfo
  {volume} {113}},\ \bibinfo {pages} {203002} (\bibinfo {year}
  {2014})}\BibitemShut {NoStop}%
\bibitem [{\citenamefont {Vukics}\ \emph {et~al.}(2007)\citenamefont {Vukics},
  \citenamefont {Maschler},\ and\ \citenamefont {Ritsch}}]{Vukics2007a}%
  \BibitemOpen
  \bibfield  {author} {\bibinfo {author} {\bibfnamefont {A.}~\bibnamefont
  {Vukics}}, \bibinfo {author} {\bibfnamefont {C.}~\bibnamefont {Maschler}}, \
  and\ \bibinfo {author} {\bibfnamefont {H.}~\bibnamefont {Ritsch}},\ }\href
  {http://dx.doi.org/10.1088/1367-2630/9/8/255} {\bibfield  {journal} {\bibinfo
   {journal} {{New J. Phys.}}\ }\textbf {\bibinfo {volume} {9}},\ \bibinfo
  {pages} {255} (\bibinfo {year} {2007})}\BibitemShut {NoStop}%
\bibitem [{\citenamefont {Fern\'{a}ndez-Vidal}\ \emph
  {et~al.}(2010)\citenamefont {Fern\'{a}ndez-Vidal}, \citenamefont {{De
  Chiara}}, \citenamefont {Larson},\ and\ \citenamefont
  {Morigi}}]{Fernandez2010quantum}%
  \BibitemOpen
  \bibfield  {author} {\bibinfo {author} {\bibfnamefont {S.}~\bibnamefont
  {Fern\'{a}ndez-Vidal}}, \bibinfo {author} {\bibfnamefont {G.}~\bibnamefont
  {{De Chiara}}}, \bibinfo {author} {\bibfnamefont {J.}~\bibnamefont {Larson}},
  \ and\ \bibinfo {author} {\bibfnamefont {G.}~\bibnamefont {Morigi}},\ }\href
  {\doibase 10.1103/PhysRevA.81.043407} {\bibfield  {journal} {\bibinfo
  {journal} {{Phys. Rev. A}}\ }\textbf {\bibinfo {volume} {81}},\ \bibinfo
  {pages} {043407} (\bibinfo {year} {2010})}\BibitemShut {NoStop}%
\bibitem [{\citenamefont {K\'{o}nya}\ \emph {et~al.}(2011)\citenamefont
  {K\'{o}nya}, \citenamefont {Szirmai},\ and\ \citenamefont
  {Domokos}}]{Konya2011}%
  \BibitemOpen
  \bibfield  {author} {\bibinfo {author} {\bibfnamefont {G.}~\bibnamefont
  {K\'{o}nya}}, \bibinfo {author} {\bibfnamefont {G.}~\bibnamefont {Szirmai}},
  \ and\ \bibinfo {author} {\bibfnamefont {P.}~\bibnamefont {Domokos}},\ }\href
  {http://www.springerlink.com/content/031278m3j2614511} {\bibfield  {journal}
  {\bibinfo  {journal} {{Eur. Phys. J. D}}\ }\textbf {\bibinfo {volume} {65}},\
  \bibinfo {pages} {33} (\bibinfo {year} {2011})}\BibitemShut {NoStop}%
\bibitem [{\citenamefont {Habibian}\ \emph {et~al.}(2013)\citenamefont
  {Habibian}, \citenamefont {Winter}, \citenamefont {Paganelli}, \citenamefont
  {Rieger},\ and\ \citenamefont {Morigi}}]{Habibian2013bose}%
  \BibitemOpen
  \bibfield  {author} {\bibinfo {author} {\bibfnamefont {H.}~\bibnamefont
  {Habibian}}, \bibinfo {author} {\bibfnamefont {A.}~\bibnamefont {Winter}},
  \bibinfo {author} {\bibfnamefont {S.}~\bibnamefont {Paganelli}}, \bibinfo
  {author} {\bibfnamefont {H.}~\bibnamefont {Rieger}}, \ and\ \bibinfo {author}
  {\bibfnamefont {G.}~\bibnamefont {Morigi}},\ }\href {\doibase
  10.1103/PhysRevLett.110.075304} {\bibfield  {journal} {\bibinfo  {journal}
  {{Phys. Rev. Lett.}}\ }\textbf {\bibinfo {volume} {110}},\ \bibinfo {pages}
  {075304} (\bibinfo {year} {2013})}\BibitemShut {NoStop}%
\bibitem [{\citenamefont {Li}\ \emph {et~al.}(2013)\citenamefont {Li},
  \citenamefont {He},\ and\ \citenamefont {Hofstetter}}]{Li2013Lattice}%
  \BibitemOpen
  \bibfield  {author} {\bibinfo {author} {\bibfnamefont {Y.}~\bibnamefont
  {Li}}, \bibinfo {author} {\bibfnamefont {L.}~\bibnamefont {He}}, \ and\
  \bibinfo {author} {\bibfnamefont {W.}~\bibnamefont {Hofstetter}},\ }\href
  {\doibase 10.1103/PhysRevA.87.051604} {\bibfield  {journal} {\bibinfo
  {journal} {{Phys. Rev. A}}\ }\textbf {\bibinfo {volume} {87}},\ \bibinfo
  {pages} {051604} (\bibinfo {year} {2013})}\BibitemShut {NoStop}%
\bibitem [{\citenamefont {Piazza}\ \emph {et~al.}(2013)\citenamefont {Piazza},
  \citenamefont {Strack},\ and\ \citenamefont {Zwerger}}]{Piazza2013bose}%
  \BibitemOpen
  \bibfield  {author} {\bibinfo {author} {\bibfnamefont {F.}~\bibnamefont
  {Piazza}}, \bibinfo {author} {\bibfnamefont {P.}~\bibnamefont {Strack}}, \
  and\ \bibinfo {author} {\bibfnamefont {W.}~\bibnamefont {Zwerger}},\ }\href
  {\doibase 10.1016/j.aop.2013.08.015} {\bibfield  {journal} {\bibinfo
  {journal} {{Ann. Phys.}}\ }\textbf {\bibinfo {volume} {339}},\ \bibinfo
  {pages} {135} (\bibinfo {year} {2013})}\BibitemShut {NoStop}%
\bibitem [{\citenamefont {Bakhtiari}\ \emph {et~al.}(2015)\citenamefont
  {Bakhtiari}, \citenamefont {Hemmerich}, \citenamefont {Ritsch},\ and\
  \citenamefont {Thorwart}}]{Bakhtiari2015Nonequilibrium}%
  \BibitemOpen
  \bibfield  {author} {\bibinfo {author} {\bibfnamefont {M.~R.}\ \bibnamefont
  {Bakhtiari}}, \bibinfo {author} {\bibfnamefont {A.}~\bibnamefont
  {Hemmerich}}, \bibinfo {author} {\bibfnamefont {H.}~\bibnamefont {Ritsch}}, \
  and\ \bibinfo {author} {\bibfnamefont {M.}~\bibnamefont {Thorwart}},\ }\href
  {\doibase 10.1103/PhysRevLett.114.123601} {\bibfield  {journal} {\bibinfo
  {journal} {{Phys. Rev. Lett.}}\ }\textbf {\bibinfo {volume} {114}},\ \bibinfo
  {pages} {123601} (\bibinfo {year} {2015})}\BibitemShut {NoStop}%
\bibitem [{\citenamefont {Black}\ \emph {et~al.}(2003)\citenamefont {Black},
  \citenamefont {Chan},\ and\ \citenamefont {{Vuleti\ifmmode \acute{c}\else
  \'{c}\fi{}}}}]{Black2003observation}%
  \BibitemOpen
  \bibfield  {author} {\bibinfo {author} {\bibfnamefont {A.~T.}\ \bibnamefont
  {Black}}, \bibinfo {author} {\bibfnamefont {H.~W.}\ \bibnamefont {Chan}}, \
  and\ \bibinfo {author} {\bibfnamefont {V.}~\bibnamefont {{Vuleti\ifmmode
  \acute{c}\else \'{c}\fi{}}}},\ }\href {\doibase
  10.1103/PhysRevLett.91.203001} {\bibfield  {journal} {\bibinfo  {journal}
  {{Phys. Rev. Lett.}}\ }\textbf {\bibinfo {volume} {91}},\ \bibinfo {pages}
  {203001} (\bibinfo {year} {2003})}\BibitemShut {NoStop}%
\bibitem [{\citenamefont {Arnold}\ \emph {et~al.}(2012)\citenamefont {Arnold},
  \citenamefont {Baden},\ and\ \citenamefont {Barrett}}]{Arnold2012self}%
  \BibitemOpen
  \bibfield  {author} {\bibinfo {author} {\bibfnamefont {K.~J.}\ \bibnamefont
  {Arnold}}, \bibinfo {author} {\bibfnamefont {M.~P.}\ \bibnamefont {Baden}}, \
  and\ \bibinfo {author} {\bibfnamefont {M.~D.}\ \bibnamefont {Barrett}},\
  }\href {\doibase 10.1103/PhysRevLett.109.153002} {\bibfield  {journal}
  {\bibinfo  {journal} {{Phys. Rev. Lett.}}\ }\textbf {\bibinfo {volume}
  {109}},\ \bibinfo {pages} {153002} (\bibinfo {year} {2012})}\BibitemShut
  {NoStop}%
\bibitem [{\citenamefont {Baumann}\ \emph {et~al.}(2010)\citenamefont
  {Baumann}, \citenamefont {Guerlin}, \citenamefont {Brennecke},\ and\
  \citenamefont {Esslinger}}]{Baumann2010dicke}%
  \BibitemOpen
  \bibfield  {author} {\bibinfo {author} {\bibfnamefont {K.}~\bibnamefont
  {Baumann}}, \bibinfo {author} {\bibfnamefont {C.}~\bibnamefont {Guerlin}},
  \bibinfo {author} {\bibfnamefont {F.}~\bibnamefont {Brennecke}}, \ and\
  \bibinfo {author} {\bibfnamefont {T.}~\bibnamefont {Esslinger}},\ }\href
  {\doibase 10.1038/nature09009} {\bibfield  {journal} {\bibinfo  {journal}
  {Nature}\ }\textbf {\bibinfo {volume} {464}},\ \bibinfo {pages} {1301}
  (\bibinfo {year} {2010})}\BibitemShut {NoStop}%
\bibitem [{\citenamefont {Mottl}\ \emph {et~al.}(2012)\citenamefont {Mottl},
  \citenamefont {Brennecke}, \citenamefont {Baumann}, \citenamefont {Landig},
  \citenamefont {Donner},\ and\ \citenamefont {Esslinger}}]{Mottl2012Roton}%
  \BibitemOpen
  \bibfield  {author} {\bibinfo {author} {\bibfnamefont {R.}~\bibnamefont
  {Mottl}}, \bibinfo {author} {\bibfnamefont {F.}~\bibnamefont {Brennecke}},
  \bibinfo {author} {\bibfnamefont {K.}~\bibnamefont {Baumann}}, \bibinfo
  {author} {\bibfnamefont {R.}~\bibnamefont {Landig}}, \bibinfo {author}
  {\bibfnamefont {T.}~\bibnamefont {Donner}}, \ and\ \bibinfo {author}
  {\bibfnamefont {T.}~\bibnamefont {Esslinger}},\ }\href {\doibase
  10.1126/science.1220314} {\bibfield  {journal} {\bibinfo  {journal}
  {Science}\ }\textbf {\bibinfo {volume} {336}},\ \bibinfo {pages} {1570}
  (\bibinfo {year} {2012})}\BibitemShut {NoStop}%
\bibitem [{\citenamefont {Ke{\ss}ler}\ \emph {et~al.}(2014)\citenamefont
  {Ke{\ss}ler}, \citenamefont {Klinder}, \citenamefont {Wolke},\ and\
  \citenamefont {Hemmerich}}]{Kessler2014Steering}%
  \BibitemOpen
  \bibfield  {author} {\bibinfo {author} {\bibfnamefont {H.}~\bibnamefont
  {Ke{\ss}ler}}, \bibinfo {author} {\bibfnamefont {J.}~\bibnamefont {Klinder}},
  \bibinfo {author} {\bibfnamefont {M.}~\bibnamefont {Wolke}}, \ and\ \bibinfo
  {author} {\bibfnamefont {A.}~\bibnamefont {Hemmerich}},\ }\href {\doibase
  10.1103/PhysRevLett.113.070404} {\bibfield  {journal} {\bibinfo  {journal}
  {{Phys. Rev. Lett.}}\ }\textbf {\bibinfo {volume} {113}},\ \bibinfo {pages}
  {070404} (\bibinfo {year} {2014})}\BibitemShut {NoStop}%
\bibitem [{\citenamefont {Dicke}(1954)}]{Dicke1954coherence}%
  \BibitemOpen
  \bibfield  {author} {\bibinfo {author} {\bibfnamefont {R.~H.}\ \bibnamefont
  {Dicke}},\ }\href {\doibase 10.1103/PhysRev.93.99} {\bibfield  {journal}
  {\bibinfo  {journal} {Phys. Rev.}\ }\textbf {\bibinfo {volume} {93}},\
  \bibinfo {pages} {99} (\bibinfo {year} {1954})}\BibitemShut {NoStop}%
\bibitem [{\citenamefont {Hepp}\ and\ \citenamefont
  {Lieb}(1973)}]{Hepp1973superradiant}%
  \BibitemOpen
  \bibfield  {author} {\bibinfo {author} {\bibfnamefont {K.}~\bibnamefont
  {Hepp}}\ and\ \bibinfo {author} {\bibfnamefont {E.~H.}\ \bibnamefont
  {Lieb}},\ }\href {\doibase 10.1016/0003-4916(73)90039-0} {\bibfield
  {journal} {\bibinfo  {journal} {{Ann. Phys.}}\ }\textbf {\bibinfo {volume}
  {76}},\ \bibinfo {pages} {360} (\bibinfo {year} {1973})}\BibitemShut
  {NoStop}%
\bibitem [{\citenamefont {Wang}\ and\ \citenamefont
  {Hioe}(1973)}]{Hioe1973Phase}%
  \BibitemOpen
  \bibfield  {author} {\bibinfo {author} {\bibfnamefont {Y.~K.}\ \bibnamefont
  {Wang}}\ and\ \bibinfo {author} {\bibfnamefont {F.~T.}\ \bibnamefont
  {Hioe}},\ }\href {\doibase 10.1103/PhysRevA.7.831} {\bibfield  {journal}
  {\bibinfo  {journal} {{Phys. Rev. A}}\ }\textbf {\bibinfo {volume} {7}},\
  \bibinfo {pages} {831} (\bibinfo {year} {1973})}\BibitemShut {NoStop}%
\bibitem [{\citenamefont {Piazza}\ and\ \citenamefont
  {Strack}(2014{\natexlab{a}})}]{Piazza2014Umklapp}%
  \BibitemOpen
  \bibfield  {author} {\bibinfo {author} {\bibfnamefont {F.}~\bibnamefont
  {Piazza}}\ and\ \bibinfo {author} {\bibfnamefont {P.}~\bibnamefont
  {Strack}},\ }\href {\doibase 10.1103/PhysRevLett.112.143003} {\bibfield
  {journal} {\bibinfo  {journal} {{Phys. Rev. Lett.}}\ }\textbf {\bibinfo
  {volume} {112}},\ \bibinfo {pages} {143003} (\bibinfo {year}
  {2014}{\natexlab{a}})}\BibitemShut {NoStop}%
\bibitem [{\citenamefont {Keeling}\ \emph {et~al.}(2014)\citenamefont
  {Keeling}, \citenamefont {Bhaseen},\ and\ \citenamefont
  {Simons}}]{Keeling2014Fermionic}%
  \BibitemOpen
  \bibfield  {author} {\bibinfo {author} {\bibfnamefont {J.}~\bibnamefont
  {Keeling}}, \bibinfo {author} {\bibfnamefont {J.}~\bibnamefont {Bhaseen}}, \
  and\ \bibinfo {author} {\bibfnamefont {B.}~\bibnamefont {Simons}},\ }\href
  {\doibase 10.1103/PhysRevLett.112.143002} {\bibfield  {journal} {\bibinfo
  {journal} {{Phys. Rev. Lett.}}\ }\textbf {\bibinfo {volume} {112}},\ \bibinfo
  {pages} {143002} (\bibinfo {year} {2014})}\BibitemShut {NoStop}%
\bibitem [{\citenamefont {Chen}\ \emph {et~al.}(2014)\citenamefont {Chen},
  \citenamefont {Yu},\ and\ \citenamefont {Zhai}}]{Chen2014Superradiance}%
  \BibitemOpen
  \bibfield  {author} {\bibinfo {author} {\bibfnamefont {Y.}~\bibnamefont
  {Chen}}, \bibinfo {author} {\bibfnamefont {Z.}~\bibnamefont {Yu}}, \ and\
  \bibinfo {author} {\bibfnamefont {H.}~\bibnamefont {Zhai}},\ }\href
  {http://journals.aps.org/prl/abstract/10.1103/PhysRevLett.112.143004}
  {\bibfield  {journal} {\bibinfo  {journal} {{Phys. Rev. Lett.}}\ }\textbf
  {\bibinfo {volume} {112}},\ \bibinfo {pages} {143004} (\bibinfo {year}
  {2014})}\BibitemShut {NoStop}%
\bibitem [{\citenamefont {Brennecke}\ \emph {et~al.}(2013)\citenamefont
  {Brennecke}, \citenamefont {Mottl}, \citenamefont {Baumann}, \citenamefont
  {Landig}, \citenamefont {Donner},\ and\ \citenamefont
  {Esslinger}}]{Brennecke2013real}%
  \BibitemOpen
  \bibfield  {author} {\bibinfo {author} {\bibfnamefont {F.}~\bibnamefont
  {Brennecke}}, \bibinfo {author} {\bibfnamefont {R.}~\bibnamefont {Mottl}},
  \bibinfo {author} {\bibfnamefont {K.}~\bibnamefont {Baumann}}, \bibinfo
  {author} {\bibfnamefont {R.}~\bibnamefont {Landig}}, \bibinfo {author}
  {\bibfnamefont {T.}~\bibnamefont {Donner}}, \ and\ \bibinfo {author}
  {\bibfnamefont {T.}~\bibnamefont {Esslinger}},\ }\href {\doibase
  10.1073/pnas.1306993110} {\bibfield  {journal} {\bibinfo  {journal} {Proc.
  Natl. Acad. Sci. U.S.A.}\ }\textbf {\bibinfo {volume} {110}},\ \bibinfo
  {pages} {11763} (\bibinfo {year} {2013})}\BibitemShut {NoStop}%
\bibitem [{\citenamefont {Klinder}\ \emph {et~al.}(2015)\citenamefont
  {Klinder}, \citenamefont {Ke{\ss}ler}, \citenamefont {Wolke}, \citenamefont
  {Mathey},\ and\ \citenamefont {Hemmerich}}]{Klinder2015Dynamical}%
  \BibitemOpen
  \bibfield  {author} {\bibinfo {author} {\bibfnamefont {J.}~\bibnamefont
  {Klinder}}, \bibinfo {author} {\bibfnamefont {H.}~\bibnamefont {Ke{\ss}ler}},
  \bibinfo {author} {\bibfnamefont {M.}~\bibnamefont {Wolke}}, \bibinfo
  {author} {\bibfnamefont {L.}~\bibnamefont {Mathey}}, \ and\ \bibinfo {author}
  {\bibfnamefont {A.}~\bibnamefont {Hemmerich}},\ }\href
  {http://www.pnas.org/content/112/11/3290} {\bibfield  {journal} {\bibinfo
  {journal} {Proc. Natl. Acad. Sci. U.S.A.}\ }\textbf {\bibinfo {volume}
  {112}},\ \bibinfo {pages} {3290} (\bibinfo {year} {2015})}\BibitemShut
  {NoStop}%
\bibitem [{\citenamefont {K\'{o}nya}\ \emph {et~al.}(2012)\citenamefont
  {K\'{o}nya}, \citenamefont {Nagy}, \citenamefont {Szirmai},\ and\
  \citenamefont {Domokos}}]{Konya2012finite}%
  \BibitemOpen
  \bibfield  {author} {\bibinfo {author} {\bibfnamefont {G.}~\bibnamefont
  {K\'{o}nya}}, \bibinfo {author} {\bibfnamefont {D.}~\bibnamefont {Nagy}},
  \bibinfo {author} {\bibfnamefont {G.}~\bibnamefont {Szirmai}}, \ and\
  \bibinfo {author} {\bibfnamefont {P.}~\bibnamefont {Domokos}},\ }\href
  {\doibase 10.1103/PhysRevA.86.013641} {\bibfield  {journal} {\bibinfo
  {journal} {{Phys. Rev. A}}\ }\textbf {\bibinfo {volume} {86}},\ \bibinfo
  {pages} {013641} (\bibinfo {year} {2012})}\BibitemShut {NoStop}%
\bibitem [{\citenamefont {Piazza}\ and\ \citenamefont
  {Strack}(2014{\natexlab{b}})}]{Piazza2014Quantum}%
  \BibitemOpen
  \bibfield  {author} {\bibinfo {author} {\bibfnamefont {F.}~\bibnamefont
  {Piazza}}\ and\ \bibinfo {author} {\bibfnamefont {P.}~\bibnamefont
  {Strack}},\ }\href {\doibase 10.1103/PhysRevA.90.043823} {\bibfield
  {journal} {\bibinfo  {journal} {{Phys. Rev. A}}\ }\textbf {\bibinfo {volume}
  {90}},\ \bibinfo {pages} {043823} (\bibinfo {year}
  {2014}{\natexlab{b}})}\BibitemShut {NoStop}%
\bibitem [{\citenamefont {Niedenzu}\ \emph {et~al.}(2012)\citenamefont
  {Niedenzu}, \citenamefont {Sandner}, \citenamefont {Genes},\ and\
  \citenamefont {Ritsch}}]{Niedenzu2012}%
  \BibitemOpen
  \bibfield  {author} {\bibinfo {author} {\bibfnamefont {W.}~\bibnamefont
  {Niedenzu}}, \bibinfo {author} {\bibfnamefont {R.~M.}\ \bibnamefont
  {Sandner}}, \bibinfo {author} {\bibfnamefont {C.}~\bibnamefont {Genes}}, \
  and\ \bibinfo {author} {\bibfnamefont {H.}~\bibnamefont {Ritsch}},\ }\href
  {http://stacks.iop.org/0953-4075/45/i=24/a=245501} {\bibfield  {journal}
  {\bibinfo  {journal} {{J. Phys. B}}\ }\textbf {\bibinfo {volume} {45}},\
  \bibinfo {pages} {245501} (\bibinfo {year} {2012})}\BibitemShut {NoStop}%
\bibitem [{\citenamefont {Gardiner}\ and\ \citenamefont
  {Zoller}(2000)}]{Gardiner2000}%
  \BibitemOpen
  \bibfield  {author} {\bibinfo {author} {\bibfnamefont {C.~W.}\ \bibnamefont
  {Gardiner}}\ and\ \bibinfo {author} {\bibfnamefont {P.}~\bibnamefont
  {Zoller}},\ }\href@noop {} {\emph {\bibinfo {title} {{Quantum Noise}}}},\
  \bibinfo {edition} {2nd}\ ed.\ (\bibinfo  {publisher} {Springer},\ \bibinfo
  {year} {2000})\BibitemShut {NoStop}%
\bibitem [{\citenamefont {Maschler}\ \emph {et~al.}(2007)\citenamefont
  {Maschler}, \citenamefont {Ritsch}, \citenamefont {Vukics},\ and\
  \citenamefont {Domokos}}]{Maschler2007Entanglement}%
  \BibitemOpen
  \bibfield  {author} {\bibinfo {author} {\bibfnamefont {C.}~\bibnamefont
  {Maschler}}, \bibinfo {author} {\bibfnamefont {H.}~\bibnamefont {Ritsch}},
  \bibinfo {author} {\bibfnamefont {A.}~\bibnamefont {Vukics}}, \ and\ \bibinfo
  {author} {\bibfnamefont {P.}~\bibnamefont {Domokos}},\ }\href
  {http://www.sciencedirect.com/science/article/pii/S0030401807000958}
  {\bibfield  {journal} {\bibinfo  {journal} {Opt. Commun.}\ }\textbf {\bibinfo
  {volume} {273}},\ \bibinfo {pages} {446} (\bibinfo {year}
  {2007})}\BibitemShut {NoStop}%
\bibitem [{\citenamefont {Walls}\ and\ \citenamefont
  {Milburn}(1994)}]{wallsbook}%
  \BibitemOpen
  \bibfield  {author} {\bibinfo {author} {\bibfnamefont {D.~F.}\ \bibnamefont
  {Walls}}\ and\ \bibinfo {author} {\bibfnamefont {G.~J.}\ \bibnamefont
  {Milburn}},\ }\href@noop {} {\emph {\bibinfo {title} {{Quantum Optics}}}},\
  \bibinfo {edition} {1st}\ ed.\ (\bibinfo  {publisher} {Springer-Verlag},\
  \bibinfo {address} {Berlin},\ \bibinfo {year} {1994})\BibitemShut {NoStop}%
\bibitem [{\citenamefont {Landig}\ \emph {et~al.}(2015)\citenamefont {Landig},
  \citenamefont {Brennecke}, \citenamefont {Mottl}, \citenamefont {Donner},\
  and\ \citenamefont {Esslinger}}]{Landig2015Measuring}%
  \BibitemOpen
  \bibfield  {author} {\bibinfo {author} {\bibfnamefont {R.}~\bibnamefont
  {Landig}}, \bibinfo {author} {\bibfnamefont {F.}~\bibnamefont {Brennecke}},
  \bibinfo {author} {\bibfnamefont {R.}~\bibnamefont {Mottl}}, \bibinfo
  {author} {\bibfnamefont {T.}~\bibnamefont {Donner}}, \ and\ \bibinfo {author}
  {\bibfnamefont {T.}~\bibnamefont {Esslinger}},\ }\href
  {http://dx.doi.org/10.1038/ncomms8046} {\bibfield  {journal} {\bibinfo
  {journal} {Nat. Commun.}\ }\textbf {\bibinfo {volume} {6}} (\bibinfo {year}
  {2015})}\BibitemShut {NoStop}%
\bibitem [{\citenamefont {Nagy}\ \emph {et~al.}(2008)\citenamefont {Nagy},
  \citenamefont {Szirmai},\ and\ \citenamefont {Domokos}}]{Nagy2008self}%
  \BibitemOpen
  \bibfield  {author} {\bibinfo {author} {\bibfnamefont {D.}~\bibnamefont
  {Nagy}}, \bibinfo {author} {\bibfnamefont {G.}~\bibnamefont {Szirmai}}, \
  and\ \bibinfo {author} {\bibfnamefont {P.}~\bibnamefont {Domokos}},\ }\href
  {\doibase 10.1140/epjd/e2008-00074-6} {\bibfield  {journal} {\bibinfo
  {journal} {{Eur. Phys. J. D}}\ }\textbf {\bibinfo {volume} {48}},\ \bibinfo
  {pages} {127} (\bibinfo {year} {2008})}\BibitemShut {NoStop}%
\bibitem [{\citenamefont {Nagy}\ \emph {et~al.}(2010)\citenamefont {Nagy},
  \citenamefont {K\'{o}nya}, \citenamefont {Szirmai},\ and\ \citenamefont
  {Domokos}}]{Nagy2010Dicke}%
  \BibitemOpen
  \bibfield  {author} {\bibinfo {author} {\bibfnamefont {D.}~\bibnamefont
  {Nagy}}, \bibinfo {author} {\bibfnamefont {G.}~\bibnamefont {K\'{o}nya}},
  \bibinfo {author} {\bibfnamefont {G.}~\bibnamefont {Szirmai}}, \ and\
  \bibinfo {author} {\bibfnamefont {P.}~\bibnamefont {Domokos}},\ }\href
  {\doibase 10.1103/PhysRevLett.104.130401} {\bibfield  {journal} {\bibinfo
  {journal} {Phys. Rev. Lett.}\ }\textbf {\bibinfo {volume} {104}},\ \bibinfo
  {pages} {130401} (\bibinfo {year} {2010})}\BibitemShut {NoStop}%
\bibitem [{\citenamefont {Yang}\ \emph {et~al.}(2014)\citenamefont {Yang},
  \citenamefont {Al-Amri},\ and\ \citenamefont {Zubairy}}]{Zubairy2014Dicke}%
  \BibitemOpen
  \bibfield  {author} {\bibinfo {author} {\bibfnamefont {S.}~\bibnamefont
  {Yang}}, \bibinfo {author} {\bibfnamefont {M.}~\bibnamefont {Al-Amri}}, \
  and\ \bibinfo {author} {\bibfnamefont {M.~S.}\ \bibnamefont {Zubairy}},\
  }\href {http://stacks.iop.org/0953-4075/47/i=13/a=135503} {\bibfield
  {journal} {\bibinfo  {journal} {{J. Phys. B}}\ }\textbf {\bibinfo {volume}
  {47}},\ \bibinfo {pages} {135503} (\bibinfo {year} {2014})}\BibitemShut
  {NoStop}%
\end{thebibliography}

\end{document}